\documentclass[useAMS,usenatbib]{mn2e}

\usepackage{amsmath}
\usepackage{amssymb}
\usepackage{graphicx}
\usepackage{dcolumn}
\usepackage{bm}
\usepackage{ulem}

\title[Breaking properties of face-centered crystals]
{Elastic and breaking properties of epitaxial face-centered 
crystals in neutron star crusts and white dwarf cores}
\author[D. A. Baiko]{D. A. Baiko\thanks{E-mail:baiko@astro.ioffe.ru} \\
Ioffe Institute, Politekhnicheskaya 26, 194021 Saint Petersburg, Russia}

\begin{document}

\date{Accepted; Received ; in original form}

\pagerange{\pageref{firstpage}--\pageref{lastpage}} \pubyear{2014}

\maketitle

\label{firstpage}

\begin{abstract}
Crystallization of dense matter in neutron star crusts and white dwarf 
cores may be similar to epitaxial crystal growth in terrestrial 
laboratories. However in stellar crystals, the spacing between 
horizontal planes has to gradually increase with the outward movement of
the crystallization front, tracing decrease of the electron density.
This process produces Coulomb crystals with stretched rather than
cubic elementary cells. We extend the analysis of the elastic and
breaking properties of such crystals to the face-centered (fc)
lattice. Shear deformations orthogonal to the stretch direction have 
been studied for 22 crystallographic shear planes. A common property 
for all these planes is a reduction and eventual nulling of the breaking 
shear strain with deviation from the unstretched configuration. The 
effective shear moduli for deformations orthogonal to the stretch 
direction have been calculated. It is possible that the epitaxial 
crystallization in compact stars results in a formation of large-scale 
crystallites or, at least, in growth of the whole crystallization front 
perpendicular to particular crystallographic planes. For fc structure 
growth orthogonal to the $\{111\}$ planes, we expect that, at any 
density, $\sim 5\%$ ($\sim 0.5\%$) of crystallite height is occupied by 
layers one (two) orders of magnitude weaker than the bulk of the 
crystallite. This may be important for realistic modeling of crustquakes 
on neutron stars. 
\end{abstract}




\section{Introduction}
Some of the best known works of Kenneth Golden and Gabor Kalman 
\citep*[][]{CKG82,KG90,GKW92,GK00,DKG02} have been devoted to theoretical studies of one-component 
plasma (OCP). The OCP is a system of identical point charges (hereafter 
ions) immersed into incompressible uniform charge-neutralizing 
background (hereafter electrons). This system is of fundamental 
importance for plasma physics and condensed matter physics as one of 
the simplest models in which effects of strong non-ideality of ions as 
well as ion quantum effects can be studied by various theoretical 
methods. The OCP is also highly relevant for astrophysics of compact 
stars, white dwarfs (WD) and neutron stars (NS) \citep*[e.g.][]{HPY07}. In these objects, under 
the action of enormous gravity, matter is compressed to extremely high 
densities (up to $\sim 10^{10}$ g/cc for WD and $\sim 10^{14}$ g/cc for 
NS crust) which are practically unreachable in Earth laboratories. At 
such densities, atoms are fully ionized whereas electrons are strongly 
degenerate. Electrons form the charge-neutralizing background which, in 
many problems, can be treated as constant and uniform \citep[e.g.][]{HPY07}.    

One of the more interesting features of the OCP in the compact stars is 
its inevitable crystallization in the course of stellar cooling \citep[e.g.][]{VH68}. The 
main difference between crystallized and liquid forms of the OCP is the
property of the former to support finite stretch and shear deformations,
to yield if a critical deformation is exceeded, and to possess a 
spectrum of transverse acoustic phonon modes related to these phenomena.   
Long-lived shear eigenmodes are also known in the liquid OCP under 
conditions of strong coupling \citep[][]{HMP75,GKW92}. There is an 
intimate connection between eigenmodes of the liquid OCP and phonon
modes in its crystallized state \citep[e.g.][]{GKW92,O+13}.

Elastic and breaking properties of the crystallized OCP are very 
important for astrophysics of compact stars. First of all, they are
crucial for theoretical studies of magnetars, NS with extremely strong 
magnetic field, in excess of $\sim 10^{14}$ G. In these objects, elastic
and breaking properties of NS crust are used for building models of 
crustquakes and for an analysis of spectra of quasi-periodic 
oscillations observed in hyperflares and associated with 
torsional vibrations of the crust \citep[e.g.][]{KB17}. Breaking strain 
of NS crust also determines maximum height of ``mountains'' which the 
crust can support. The mountains, in turn, determine the mass 
quadrupole of a NS, a non-zero value of which is necessary for a 
rotating NS to emit gravitational waves \citep[][]{UCB00}.     

It is long established that the crystallization front in compact stars
moves from the deeper layers up, towards the surface \citep[e.g.][]{VH68}. This is connected
with the fact that the thermal conductivity of degenerate electrons
is very high and the stellar temperature has a less steep profile than 
that of the crystallization temperature. Crystallization in these stars 
is a near-equilibrium process which takes billions of years to complete 
in WD and hundreds-to-thousands years in NS crust. The outcome of such 
near-equilibrium bottom-up crystallization of dense matter has been 
revisited in a recent paper \citep[][hereafter Paper I]{B24}. In this 
work, it has been argued that already formed crystal layers create 
above themselves an ordered two-dimensional relief of electrostatic 
potential. The scale of variations of the potential  
(e.g. potential well depth) is significantly greater than typical ion 
kinetic energies at crystallization. The lateral extent of the wells 
is comparable with the lattice spacing, so that newly crystallizing ions 
can be easily captured by them. There is no dependence on orientation of 
impinging ions (as would be the case for covalent bonds). Vertical 
positions of the potential minima with respect to the last crystallized 
surface are controlled by the density of electrons. All these properties 
tend to facilitate two-dimensional nucleation of new crystal layers on 
top of the previous ones as opposed to three-dimensional nucleation of 
new crystallites (made of repulsing atomic nuclei) in the midst of the 
liquid. In semiconductor physics and industry, such crystal growth from 
a liquid which preserves the microscopic order of a seed is known as 
the liquid-phase epitaxy (see Paper I for an extensive list of references). 
 
Epitaxial crystal growth preserves the ion surface density, $\sigma$.
From charge neutrality, $\sigma = \eta n_{\rm e}/Z$, where $\eta$ is the
interplane spacing in the vertical direction, $n_{\rm e}$ is the 
electron density, and $Z$ is the ion charge number. Consequently, the 
epitaxial growth is accompanied by gradual vertical stretching 
(more precisely, elongation) of crystal elementary cells in response to 
pressure and electron density reduction as the crystallization front 
moves outward (Paper I). In other words, $\eta$ gradually increases while 
horizontal ion spacings remain constant. We characterize the amount of 
stretch by the stretch factor, $\xi$. The stretch factor is the ratio 
of vertical to horizontal lattice scales which would be equal to 1 in 
an unstretched lattice. Elongation corresponds to $\xi>1$. Growth of 
contracted crystals ($\xi<1$) is also possible if the front moves 
across a (relatively thin) layer where the average ion charge number 
$\langle Z \rangle$ decreases with decrease of depth (at approximately 
constant $n_{\rm e}$) under the assumption that a regular lattice still 
forms throughout (Paper I). Finally, it is known that overstretched crystals 
develop unstable phonon modes and break, which limits the vertical 
sizes of growing crystallites \citep[][]{BK17,BC18}.

\section{Methods} 
\subsection{General remarks}
In this paper, we aim to extend the results of Paper I from 
body-centered cubic (bcc) to face-centered cubic (fcc) crystals. Let us 
note that, taking into account the possibility of crystal stretching, it 
will be more accurate in some contexts to speak about general 
face-centered (fc) or body-centered (bc) lattices. In terrestrial 
experiments on crystal growth, the fcc lattice occurs much more often 
than the bcc one with growth perpendicular to $\{111\}$ planes 
particularly prominent. On the contrary, in the literature on compact 
stars, the bcc lattice is more popular due to its advantage over any 
other lattice in terms of the potential (Madelung) energy for the 
OCP of bare atomic nuclei on the constant and 
uniform neutralizing background. However, the difference between bcc 
and fcc energies for unstretched material is tiny, and once the 
possibility of crystal stretch is allowed, the preference for a bc 
lattice becomes even less solidly justified.    

We shall begin by calculating breaking strain for shear 
deformations perpendicular to a crystal stretch direction. As mentioned
in Paper I, such statement of the problem is relevant for astrophysics
of compact stars in which stretch is aligned with gravity whereas 
horizontal shear can be due to weaker forces (e.g. magnetic) which do 
not perturb the hydrostatic equilibrium. Since a 
crystal can be oriented differently and thus stretched in different 
crystallographic directions, this may
involve shear in different crystallographic planes. Within each plane, 
any azimuthal angle of the shear is possible. 

In the {\it polycrystalline} picture of stellar matter described in Paper I, 
namely, if, at any given depth, there exist randomly oriented 
crystallites stretched vertically by the same factor $\xi$, we expect 
breaking of matter to be associated with the 
crystallite with the minimum breaking shear strain. To find the latter, 
at each $\xi$ and for each plane, we minimize the breaking strain 
over the azimuthal angle and then minimize over the planes.  
In Paper I, for the bc lattice, we have performed minimization over 
3 high-symmetry planes and 4 auxiliary, less symmetric planes. For this 
work on the fc lattice, we have significantly expanded the set of 
considered configurations and have studied shear deformations in all 
crystallographic planes with Miller indices $\leq 4$. 
There are 22 such planes, which are listed in Table \ref{latparam}.

\subsection{Lattice geometry}
Let us specify a cartesian reference frame whose axes are aligned with 
the fcc lattice cube (before stretch) and one of the ions is taken as 
the origin. 
The shear planes which we consider, have intercepts with $x$, $y$, and 
$z$-axes denoted as $a$, $b$, and $c$, respectively. Based on lattice 
symmetry, we require that, if two of the intercepts are at infinity, 
then $a>0$ is finite. This is the case of crystal growth perpendicular 
to a $\{100\}$ plane. If only one of the intercepts is at infinity, then
$a$ and $b$ are finite, $a \geq b >0$, and the equality corresponds to
growth perpendicular to a $\{110\}$ plane. If all intercepts are finite,
then either $a>b>c>0$ or $a=b>0$ whereas $c>0$ can be less, greater, or
equal to $a=b$. The case, where all three intercepts
are equal, corresponds to the crystal growth perpendicular to a 
$\{111\}$ plane.
                                         
Let us define basis vectors of a new cartesian reference frame by their
coordinates in the original basis as $\bm{e}_1 = (-a,b,0)/\sqrt{s_2}$,
$\bm{e}_2=(-ab^2, -a^2b, cs_2)/\sqrt{s_2 s_4}$, and 
$\bm{e}_3=(bc,ca,ab)/\sqrt{s_4}$, where $s_2=a^2+b^2$ and 
$s_4 = a^2b^2+b^2c^2+c^2a^2$. If $c\to\infty$, $\bm{e}_2 \to (0,0,1)$ 
and $\bm{e}_3 \to (b,a,0)/\sqrt{s_2}$. If also $b\to\infty$, then 
$\bm{e}_1 \to (0,1,0)$ and, further, $\bm{e}_3 \to (1,0,0)$. In all 
cases, vector $\bm{e}_3$ coincides with the stretch direction,
whereas vectors $\bm{e}_1$ and $\bm{e}_2$ belong to the plane of the 
shear.

\begin{table*}
\centering
\caption{Lattice parameters. Column 1 shows Miller indices of a shear 
plane and the number of equivalent planes. Columns 2 and 3 are 
coordinates of direct and reciprocal basis vectors, respectively, in the 
basis described in the text.}
\resizebox{165mm}{!}{
\begin{tabular}{ccc}
\hline
\hline
     Miller index $\vert$  $N_{\rm eqv}$ 
   & direct lattice basis (units $a_{\rm l}/2$) 
   & reciprocal lattice basis (units $2 \upi /a_{\rm l}$) \\
\hline
 cube-diagonal,  $\{ 111 \}$        & $(\sqrt{2},0,0)$ & $(\sqrt{2},-\sqrt{2/3},
           -(1/\sqrt{3} + \sqrt{3/2}\, \theta_1 - \theta_2/\sqrt{2})/\xi) $ 
           \\
 8 & $(1/\sqrt{2},\sqrt{3/2},0)$ & $(0,\sqrt{8/3},
           -(1/\sqrt{3}+\sqrt{2}\,\theta_2)/\xi)$ \\
           & $(1/\sqrt{2}+\theta_1,1/\sqrt{6}+\theta_2,\xi \sqrt{4/3})$ 
           & $(0,0,\sqrt{3}/\xi)$ \\
\hline
 cube-edge,   $\{ 100 \}$       & $(2,0,0)$ & $(1,-1,-(1+\theta_1-\theta_2)/\xi)$ 
           \\
 6 & $(1,1,0)$ & $(0,2,-2 \theta_2/\xi)$ \\
           & $(1+\theta_1,\theta_2,\xi)$ & $(0,0,2/\xi)$ \\
\hline
 face-diagonal,  $\{ 110 \}$        & $(\sqrt{2},0,0)$ & $(\sqrt{2},0,
           -(\sqrt{2}+2 \theta_1)/\xi)$ 
           \\
 12 & $(0,2,0)$ & $(0,1,
           -\sqrt{2}(1+\theta_2)/\xi)$ \\
           & $(1/\sqrt{2}+\theta_1,1+\theta_2,\xi/\sqrt{2})$ 
           & $(0,0,\sqrt{8}/\xi)$  \\
\hline
  $\{ 210 \}$         & $(\sqrt{20},0,0)$ & $(1/\sqrt{5},-1,
           -(3/\sqrt{5}+\theta_1-\sqrt{5}\,\theta_2)/\xi)$ \\
 24     & $(\sqrt{5},1,0)$ & $(0,2,-\sqrt{20}\,\theta_2/\xi)$ 
           \\
           & $(3/\sqrt{5}+\theta_1,\theta_2,\xi/\sqrt{5})$ 
           & $(0,0,\sqrt{20}/\xi)$ \\
\hline
  $\{ 310 \}$         & $(\sqrt{10},0,0)$ & $(2/\sqrt{10},0,
           -(14/\sqrt{10}+2 \theta_1)/\xi)$ \\
 24     & $(0,2,0)$ & $(0,1,-\sqrt{10} (1+\theta_2)/\xi)$ 
           \\
           & $(7/\sqrt{10}+\theta_1,1+\theta_2,\xi/\sqrt{10})$ 
           & $(0,0,\sqrt{40}/\xi)$ \\
\hline
  $\{ 320 \}$         & $(\sqrt{52},0,0)$ & $(1/\sqrt{13},-1,
           -(5/\sqrt{13}+\theta_1-\sqrt{13}\,\theta_2)/\xi)$ \\
 24     & $(\sqrt{13},1,0)$ & $(0,2,-\sqrt{52}\,\theta_2/\xi)$ 
           \\
           & $(5/\sqrt{13}+\theta_1,\theta_2,\xi/\sqrt{13})$ 
           & $(0,0,\sqrt{52}/\xi)$ \\
\hline
  $\{ 211 \}$         & $(\sqrt{2},0,0)$ & $(\sqrt{2},0,
           -\sqrt{6}(1+\sqrt{2}\,\theta_1)/\xi)$ 
           \\
 24     & $(0,\sqrt{12},0)$ & $(0,1/\sqrt{3},
           -\sqrt{2}(2/\sqrt{3}+\theta_2)/\xi)$ 
           \\
           & $(1/\sqrt{2}+\theta_1,2/\sqrt{3}+\theta_2,\xi/\sqrt{6})$ 
           & $(0,0,\sqrt{24}/\xi)$ \\
\hline
  $\{ 221 \}$         & $(\sqrt{2},0,0)$ & $(\sqrt{2},0,
           -3(1+\sqrt{2}\,\theta_1)/\xi)$ 
           \\
 24     & $(0,\sqrt{18},0)$ & $(0,\sqrt{2}/3,
           -(13/3+\sqrt{2}\,\theta_2)/\xi)$ 
           \\
           & $(1/\sqrt{2}+\theta_1,13/\sqrt{18}+\theta_2,\xi/3)$ 
           & $(0,0,6/\xi)$ \\
\hline
  $\{ 311 \}$         & $(\sqrt{2},0,0)$ & $(\sqrt{2},-\sqrt{2/11},
           -(3/\sqrt{11}+\sqrt{11/2}\,\theta_1-\theta_2/\sqrt{2})/\xi)$ 
           \\
 24     & $(1/\sqrt{2},\sqrt{11/2},0)$ & $(0,\sqrt{8/11},
           -(5/\sqrt{11}+\sqrt{2}\,\theta_2)/\xi)$ 
           \\
           & $(1/\sqrt{2}+\theta_1,5/\sqrt{22}+\theta_2,2\xi/\sqrt{11})$ 
           & $(0,0,\sqrt{11}/\xi)$ \\
\hline
  $\{ 331 \}$         & $(\sqrt{2},0,0)$ & $(\sqrt{2},-\sqrt{2/19},
           (6/\sqrt{19}-\sqrt{19/2}\,\theta_1+\theta_2/\sqrt{2})/\xi)$ 
           \\
 24     & $(1/\sqrt{2},\sqrt{19/2},0)$ & $(0,\sqrt{8/19},
           -(12/\sqrt{19}+\sqrt{2}\,\theta_2)/\xi)$ 
           \\
           & $(\theta_1,\sqrt{72/19}+\theta_2,2\xi/\sqrt{19})$ 
           & $(0,0,\sqrt{19}/\xi)$ \\
\hline
  $\{ 421 \}$         & $(\sqrt{20},0,0)$ & $(1/\sqrt{5},-\sqrt{3/35},
           -\sqrt{21}(5/7+\theta_1/\sqrt{5}
           -\sqrt{3/35}\,\theta_2)/\xi)$ \\
 48     & $(3/\sqrt{5},\sqrt{21/5},0)$ & $(0,\sqrt{20/21},
           -2(17/\sqrt{21}+\sqrt{5}\,\theta_2)/\xi)$ 
           \\
           & $(6/\sqrt{5}+\theta_1,17/\sqrt{105}+\theta_2,
           \xi/\sqrt{21})$ & $(0,0,\sqrt{84}/\xi)$ \\
\hline
  $\{ 321 \}$         & $(\sqrt{20},0,0)$ & $(1/\sqrt{5},-\sqrt{18/35},
           -\sqrt{14}(6/7+\theta_1/\sqrt{5}
           -\sqrt{18/35}\,\theta_2)/\xi)$ \\
 48     & $(6/\sqrt{5},\sqrt{14/5},0)$ & $(0,\sqrt{10/7},
           -\sqrt{2}(11/\sqrt{7}+\sqrt{10}\,\theta_2)/\xi)$ 
           \\
           & $(9/\sqrt{5}+\theta_1,11/\sqrt{70}+\theta_2,
           \xi/\sqrt{14})$ & $(0,0,\sqrt{56}/\xi)$ \\
\hline
  $\{ 431 \}$         & $(\sqrt{10},0,0)$ & $(\sqrt{2/5},-2/\sqrt{65},
           -\sqrt{26}(19/13+\sqrt{2/5}\,\theta_1
           -2\theta_2/\sqrt{65})/\xi)$ \\
 48     & $(\sqrt{8/5},\sqrt{52/5},0)$ & $(0,\sqrt{5/13},
           -\sqrt{2}(11/\sqrt{13}+\sqrt{5}\,\theta_2)/\xi)$ 
           \\
           & $(9/\sqrt{10}+\theta_1,11/\sqrt{65}+\theta_2,
           \xi/\sqrt{26})$ & $(0,0,\sqrt{104}/\xi)$ \\
\hline
  $\{ 432 \}$         & $(\sqrt{52},0,0)$ & $(1/\sqrt{13},-19/\sqrt{13\cdot29},
           -\sqrt{29}(17/29+\theta_1/\sqrt{13}
           -19\,\theta_2/\sqrt{13\cdot29})/\xi)$ \\
 48     & $(19/\sqrt{13},\sqrt{29/13},0)$ & $(0,\sqrt{52/29},
           - 2(25/\sqrt{29}+\sqrt{13}\,\theta_2)/\xi)$ 
           \\
           & $(24/\sqrt{13}+\theta_1,25/\sqrt{13\cdot29}+\theta_2,
           \xi/\sqrt{29})$ & $(0,0,\sqrt{116}/\xi)$ \\
\hline
  $\{ 322 \}$         & $(\sqrt{2},0,0)$ & $(\sqrt{2},0,
           -\sqrt{17}(1+\sqrt{2}\,\theta_1)/\xi)$ 
           \\
 24     & $(0,\sqrt{34},0)$ & $(0,\sqrt{2/17},
           -\sqrt{2}(7/\sqrt{34}+\theta_2)/\xi)$ 
           \\
           & $(1/\sqrt{2}+\theta_1,7/\sqrt{34}+\theta_2,\xi/\sqrt{17})$ 
           & $(0,0,\sqrt{68}/\xi)$ \\
\hline
  $\{ 332 \}$         & $(\sqrt{2},0,0)$ & $(\sqrt{2},0,
           -\sqrt{22}(1+\sqrt{2}\,\theta_1)/\xi)$ 
           \\
 24     & $(0,\sqrt{44},0)$ & $(0,1/\sqrt{11},
           -\sqrt{2}(18/\sqrt{11}+\theta_2)/\xi)$ 
           \\
           & $(1/\sqrt{2}+\theta_1,18/\sqrt{11}+\theta_2,\xi/\sqrt{22})$ 
           & $(0,0,\sqrt{88}/\xi)$ \\
\hline
  $\{ 410 \}$         & $(\sqrt{68},0,0)$ & $(1/\sqrt{17},-1,
           -(13/\sqrt{17}+\theta_1-\sqrt{17}\,\theta_2)/\xi)$ 
           \\
 24     & $(\sqrt{17},1,0)$ & $(0,2,
           -\sqrt{68}\,\theta_2/\xi)$ 
           \\
           & $(13/\sqrt{17}+\theta_1,\theta_2,\xi/\sqrt{17})$ 
           & $(0,0,\sqrt{68}/\xi)$ \\
\hline
  $\{ 430 \}$         & $(10,0,0)$ & $(1/5,-1,
           -(7/5+\theta_1-5 \theta_2)/\xi)$ 
           \\
 24     & $(5,1,0)$ & $(0,2,
           -10 \theta_2/\xi)$ 
            \\
           & $(7/5+\theta_1,\theta_2,\xi/5)$ 
           & $(0,0,10/\xi)$ \\
\hline
  $\{ 411 \}$         & $(\sqrt{2},0,0)$ & $(\sqrt{2},0,
           -\sqrt{18}(1+\sqrt{2}\,\theta_1)/\xi)$ 
           \\
 24     & $(0,6,0)$ & $(0,1/3,
           -\sqrt{2}(7/3+\theta_2)/\xi)$ 
           \\
           & $(1/\sqrt{2}+\theta_1,7/3+\theta_2,\xi/\sqrt{18})$ 
           & $(0,0,\sqrt{72}/\xi)$ \\
\hline
  $\{ 441 \}$         & $(\sqrt{2},0,0)$ & $(\sqrt{2},0,
           -\sqrt{33}(1+\sqrt{2}\,\theta_1)/\xi)$ 
           \\
 24     & $(0,\sqrt{66},0)$ & $(0,\sqrt{2/33},
           -\sqrt{2}(41/\sqrt{66}+\theta_2)/\xi)$ 
           \\
           & $(1/\sqrt{2}+\theta_1,41/\sqrt{66}+\theta_2,\xi/\sqrt{33})$ 
           & $(0,0,\sqrt{132}/\xi)$ \\
\hline
  $\{ 433 \}$         & $(\sqrt{2},0,0)$ & $(\sqrt{2},0,
           -\sqrt{34}(1+\sqrt{2}\,\theta_1)/\xi)$ 
           \\
 24     & $(0,\sqrt{68},0)$ & $(0,1/\sqrt{17},
           -\sqrt{2}(5/\sqrt{17}+\theta_2)/\xi)$ 
           \\
           & $(1/\sqrt{2}+\theta_1,5/\sqrt{17}+\theta_2,\xi/\sqrt{34})$ 
           & $(0,0,\sqrt{136}/\xi)$ \\
\hline
  $\{ 443 \}$         & $(\sqrt{2},0,0)$ & $(\sqrt{2},0,
           -\sqrt{41}(1+\sqrt{2}\,\theta_1)/\xi)$ 
           \\
 24     & $(0,\sqrt{82},0)$ & $(0,\sqrt{2/41},
           -\sqrt{2}(71/\sqrt{82}+\theta_2)/\xi)$ 
           \\
           & $(1/\sqrt{2}+\theta_1,71/\sqrt{82}+\theta_2,\xi/\sqrt{41})$ 
           & $(0,0,\sqrt{164}/\xi)$ \\
\hline
\end{tabular}
}
\label{latparam}
\end{table*}

The stretched fc lattice basis vectors and the respective reciprocal 
lattice basis vectors in the basis $(\bm{e}_1, \bm{e}_2, \bm{e}_3)$ are 
given in Table \ref{latparam} for 22 variants of the shear plane. 
For non-deformed fcc lattice, one has to set 
$\xi=1$ and $\theta_1=\theta_2=0$ in these expressions. To find the 
breaking shear strain, we stretch and then shear the crystal, which is 
accomplished by a modification of the third lattice basis vector:
its $\bm{e}_3$-component is multiplied by $\xi$ for stretch, and then, 
for shear, its in-plane components are augmented by 
$\theta_1 = \theta \cos{\chi}$ and $\theta_2 = \theta \sin{\chi}$. 
In this case, $\theta$ is the amplitude of the shear and $\chi$ is its 
azimuthal angle.

\subsection{Unstable phonon modes}
Given direct and reciprocal vectors of a deformed lattice, one can 
construct its dynamic matrix which yields frequencies of the lattice 
eigenmodes (phonons). An initially stable lattice loses stability under 
a large enough deformation which is signified by the appearance of a 
phonon mode with a negative
squared frequency at some wave vector. In the problem of Coulomb crystal
breaking under an excessive stretch or shear deformation, the unstable 
mode first appears at the Brillouin zone center. Near the center, the
phonon frequencies are strongly dependent on the wave vector direction.
Strictly speaking, at the center, the phonon frequencies are non-analytic. 
With deformation increase beyond the critical magnitude, the 
zone of unstable wave vectors rapidly widens over some finite range. 
Consequently, we approximate the critical deformation by the value
which produces an unstable mode at a very small but finite wave vector.  

We scanned a dense grid of non-equivalent spherical angles of 
a phonon wave vector assuming its length equal to 
$(1/50) 2 \upi/a_{\rm l}$. For deformed fc lattices, $a_{\rm l}$ is 
defined by $n_{\rm i} a_{\rm l}^3 \xi = 4$, where $n_{\rm i}$ is the ion
density. Given a shear plane and $\chi$, at each $\xi$, we looked for 
the maximum shear deformation of a stretched crystal, at which the 
dynamic matrix determinant remained positive. We then minimized these 
maximum deformations over a dense grid of $\chi$.

\section{Results}
\subsection{Breaking shear strain of dense stretched matter}
\begin{figure*}
\begin{center}
\leavevmode
\includegraphics[bb=70 344 569 740, width=170mm]{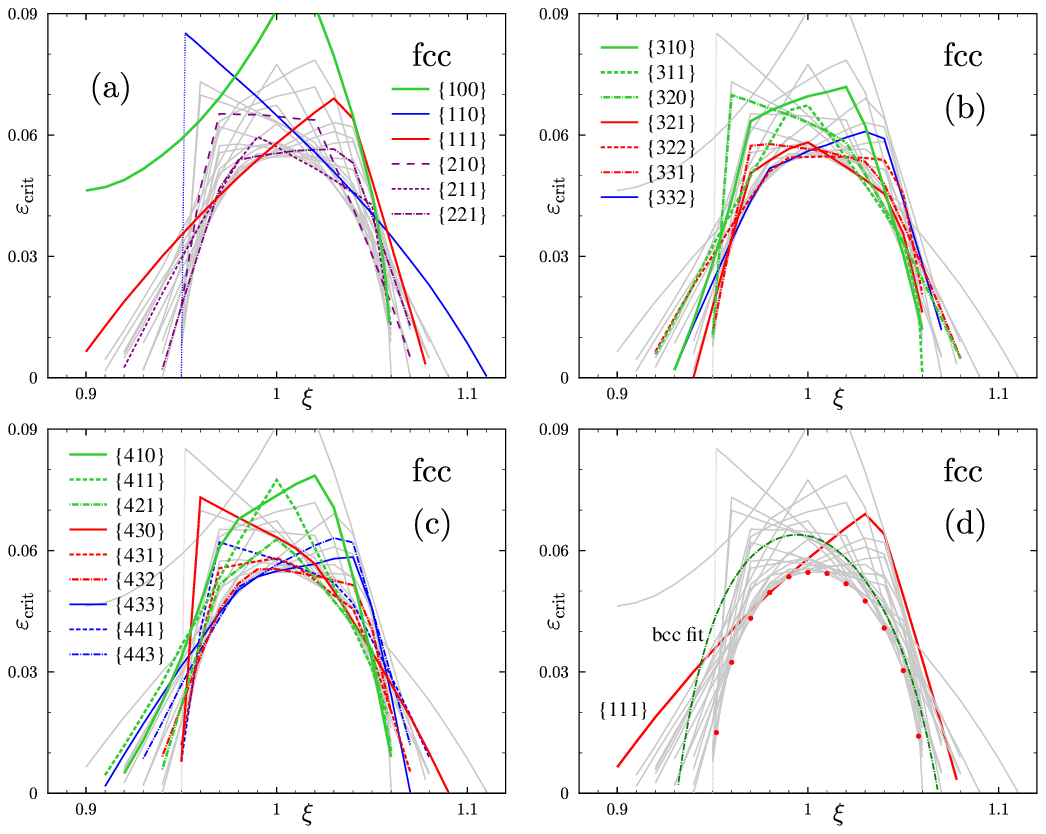}
\end{center}
\vspace{-0.2cm}
\caption[ ]{Breaking shear strain of stretched fc lattice for 22 
crystallographic shear planes with Miller indices $\leq 4$. 
All panels: light-grey lines show results for all planes. (a--c) Curves of 
various colours, types, and thicknesses display breaking shear strain for 
specific planes listed in the legends. (d) Dots show present $\varepsilon_{\rm crit}^{\min}(\xi)$,
dot-dashed curve is the fit of $\varepsilon_{\rm crit}^{\min}(\xi)$ from
Paper I for stretched bc lattice, solid red curve reproduces 
$\varepsilon_{\rm crit}(\xi)$ for growth orthogonal to the 
$\{111\}$ planes from panel a.}
\label{fbreak}
\end{figure*}

In order to represent the results, let us recall the standard strain 
tensor definition 
$u_{ij} = 0.5 (\partial u_i/\partial r_j+\partial u_j/\partial r_i)$,
where $u_i$ is the displacement, which, in this problem, we shall 
evaluate with respect to the stretched configuration; $i$ and $j$ are 
cartesian indices. Let us define
two cartesian axes, $\hat{\theta}$ and $\hat{\eta}=\bm{e}_3$, along the 
direction of the shear and along the direction of the stretch, 
respectively. The applied shear results in the appearance of  
non-zero components
$u_{\theta \eta}=u_{\eta \theta} \equiv \varepsilon/2 = \theta/(2\eta)$,
where $\varepsilon$ is the deformation parameter and $\eta$ is the 
interplane spacing. The latter is equal to the $\bm{e}_3$-component of
the third lattice basis vector (cf. Table \ref{latparam}).  
We denote the maximum deformation parameter, at which the lattice is 
still stable, minimized over $\chi$ as
$\varepsilon_{\rm crit}$ and plot it in Fig.\ \ref{fbreak} 
as a function of $\xi$, separately for each crystallographic plane 
considered. In particular, we plot $\varepsilon_{\rm crit}$ for 
crystals stretched perpendicular to crystallographic planes with 
the maximum Miller index equal to 1 or 2 (panel a), 3 (panel b), and 4 
(panel c).

In all panels of Fig.\ \ref{fbreak}, grey lines show the entire family of 
$\varepsilon_{\rm crit}(\xi)$ curves for 22 considered planes. In panel
\ref{fbreak}a, thick (green), intermediate (red), and thin (blue) curves
correspond to stretches orthogonal to $\{100\}$, $\{111\}$, and 
$\{110\}$ planes, respectively. Purple long-dashed, short-dashed, and dot-dashed
curves correspond to stretches orthogonal to $\{210\}$, $\{211\}$, and 
$\{221\}$ planes, respectively. In panels \ref{fbreak}b and c, coloured
curves of various types display $\varepsilon_{\rm crit}(\xi)$
for stretches orthogonal to planes with the maximum Miller index of 3 
and 4, respectively, as detailed in the legends.          

These results are very similar to those for the bc lattice obtained in
Paper I. A common property for all\footnote{except for $\{100\}$ planes
which can tolerate a very strong contraction from $\xi=1$ to 
$\xi=1/\sqrt{2}$, describing a structural transition from fcc to bcc 
lattice; see also \citet{BK17} and Paper I} planes is a reduction of  
breaking shear strain with deviation of the stretch factor from 1 and 
especially with its approach to the breaking limit for stretch, 
$\xi_{\rm crit}$, defined as $\varepsilon_{\rm crit}(\xi_{\rm crit})=0$. 
Shear deformations in different planes are characterized by significantly 
different breaking properties.

\begin{table}
\centering
\caption{Breaking strain. Columns 1, 2, 3, and 4 show stretch factor, 
minimized breaking shear 
strain, Miller indices of the plane in which the minimum is achieved, and 
breaking shear strain for $\{111\}$ planes, respectively.}
\begin{tabular}{cccc}
\hline
\hline
   $\xi$ & $\varepsilon_{\rm crit}^{\rm min}$ &  Miller index  & 
   $\varepsilon^{\{111\}}_{\rm crit}$ \\
\hline
  0.9     & & & 0.0065\\
  0.91    & & & 0.0127 \\
  0.92    & & & 0.0188 \\
  0.93    & & & 0.0244 \\
  0.94    & & & 0.0300 \\
  0.95    & & & 0.0352 \\
  0.952  &  0.0150    &      $\{441\}$  & \\
  0.96   &  0.0324    &      $\{221\}$  & 0.0403 \\
  0.97   &  0.0433    &      $\{443\}$  & 0.0451 \\
  0.98   &  0.0496    &      $\{111\}$  & 0.0496 \\
  0.99   &  0.0535    &      $\{443\}$  & 0.0539 \\
   1     &  0.0546    &      $\{322\}$  & 0.0580 \\
   1.01  &  0.0543    &      $\{211\}$  & 0.0618 \\
   1.02  &  0.0518    &      $\{211\}$  & 0.0656 \\
   1.03  &  0.0475    &      $\{421\}$  & 0.0690 \\
   1.04  &  0.0408    &      $\{311\}$  & 0.0641 \\
   1.05  &  0.0303    &      $\{210\}$  & 0.0488 \\
   1.058 &  0.0141    &      $\{410\}$  & \\
   1.06    & & & 0.0330 \\
   1.07    & & & 0.0168 \\
   1.072    & & & 0.0135 \\
   1.074    & & & 0.0101 \\
   1.076    & & & 0.0067 \\
   1.078    & & & 0.0034 \\
\hline
\end{tabular}
\label{epsdata}
\end{table}

For the {\it polycrystalline} scenario, in panel \ref{fbreak}d, red dots show 
minimized over 22 planes breaking shear strain. These data are given in the
second column of Table \ref{epsdata} with Miller indices of a plane, 
realizing the minimum at a given $\xi$, reported in the third column.
The minimized breaking shear strain can be fitted to rms fit accuracy 
better than 0.05\% by a simple analytic expression:
\begin{eqnarray}
    \varepsilon_{\rm crit}^{\rm min} &=& 0.0546-20.2\,(\xi^{2/3}-1.0017)^2
\nonumber \\    
              &-& 7.6\mathrm{e}6\,(\xi^{2/3}-1.0036)^6~.
\label{epsmin}
\end{eqnarray}
For comparison, dot-dashed (dark-green) curve demonstrates the analogous fit
obtained in Paper I (equation 14) for stretched bc lattice. 
Just like 
in the case of the bc lattice,
the $\varepsilon_{\rm crit}^{\rm min}(\xi)$ curve is essentially 
parabolic. 

For the unstretched fcc lattice, 
$\varepsilon_{\rm crit}^{\rm min}(1) \approx 0.055$ (to be compared with
$\approx 0.064$ for bcc). 
Breaking stretch factor for the
fc lattice is found as $\xi_{\rm crit}\approx 0.95$ and 1.06 for contractions
and elongations, respectively. Introducing stretch strain 
$\tilde{\varepsilon} \equiv \xi^{2/3}-1$, we obtain, respectively,  
$\tilde{\varepsilon}_{\rm crit}\approx -0.034$ and 0.04. This can be
compared with $|\tilde{\varepsilon}_{\rm crit}|\approx 0.04$ reported
for the bc lattice by \citet{BC18} (in reasonable agreement with 
the dot-dashed curve in Fig.\ \ref{fbreak}d).  

In general, we see that the locus of breaking strain minima of the fc
lattice lies systematically below that for the bc 
structure. We do not think that this is related to the fact that fewer 
stretch orientations have been analyzed in Paper I, because in both cases, for 
certain $\xi$, the minimum curve is bounded by critical curves for 
low-index planes.

The solid red 
curve in Fig.\ \ref{fbreak}d, the same as in Fig.\ \ref{fbreak}a, shows 
$\varepsilon_{\rm crit}(\xi)$ for stretched fc lattice, growing 
perpendicular to a $\{111\}$ plane. These are the close-packed planes
of the fcc lattice. Consequently, they are the slowest growing and they 
should, according to Bravais's rule, represent the crystallization 
front in equilibrium (see Paper I for details). Let us reiterate that
this idea is supported by the prominence of $\{111\}$ growth in 
terrestrial experiments on various fcc materials. 
In the case of near-equilibrium bottom-up 
crystallization in a NS crust or a WD core, we expect
this type of growth in the {\it macrocrystallite} scenario (Paper I). Then
the red curve rather than the minimum curve would determine the strength
of the crystallized matter. 

The data for the
red curve are given in the fourth column of Table \ref{epsdata}. To rms fit 
accuracy better than 0.01\%, the data can be analytically approximated 
as:  
\begin{equation}
    \varepsilon_{\rm crit}^{\{111\}} = \begin{cases}  
    0.157-0.099\,\xi^{-4}, & \xi < 1.0355 \\
    0.4572-0.336\,\xi^4, &  \xi > 1.0355 ~.   \end{cases}
\label{eps111}
\end{equation}
%


\subsection{Shear modulus}
If a shear deformation is not large enough to break a crystal,
it may manifest itself as a transverse shear oscillation, which
belongs to a class of stellar seismic modes. To describe dynamics of 
such modes, one needs an expression for their potential energy, which is
the crystal electrostatic energy. For weak modes, the latter is a 
quadratic function of the strain tensor. On the other hand, we can 
calculate the electrostatic 
energy increment for an infinitesimal shear of a stretched crystal, 
$\delta U$, with the aid of the lattice vectors from Table 
\ref{latparam}. Identifying further 
$(\delta U)/V = 2 \mu u^2_{\theta \eta}$,
where $V$ is the volume, we see that the quadratic dependence 
can be specified by an elastic coefficient $\mu$, which can be thus 
determined for any stretch direction, at any $\xi$, and for any 
azimuthal angle of the shear. 

\begin{figure*}
\begin{center}
\leavevmode
\includegraphics[bb=71 543 569 741, width=170mm]{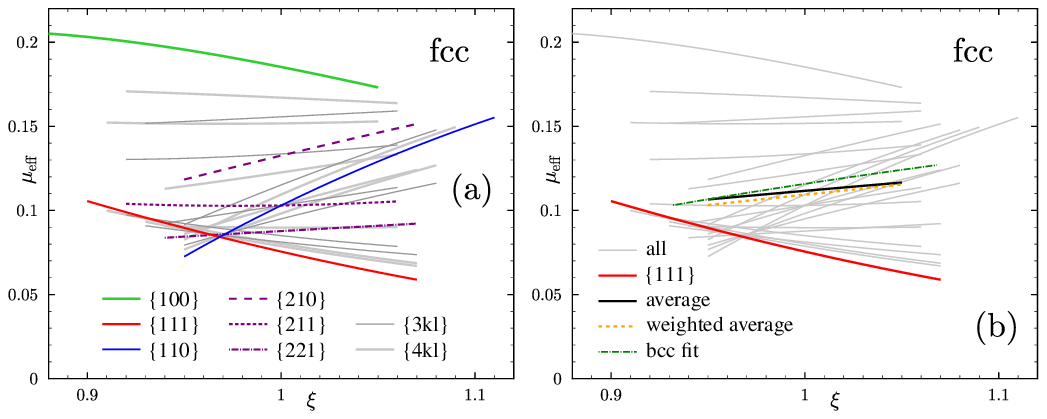}
\end{center}
\vspace{-0.2cm}
\caption[ ]{Effective shear moduli of stretched fc lattice vs. stretch
factor. (a) All considered shear planes. (b) All considered shear 
planes (solid light-grey),
$\{111\}$ planes (solid red), simple-average shear modulus (solid black),
weighted average shear modulus (dashed orange), simple-average shear
modulus of stretched bc lattice (fit from Paper I, dot-dashed darkgreen). 
}
\label{muaver}
\end{figure*}

Dependence of $\mu$ on 
the azimuthal angle turns out to be weak and, besides, there do 
not seem to be any arguments in favor of a particular azimuthal shear
direction. Thus we shall average $\mu$ over the azimuthal angle. 
This yields the effective shear modulus $\mu_{\rm eff}$, which becomes a 
function of the growth direction and the stretch factor. The quantity 
$\mu_{\rm eff}(\xi)$ is shown in Fig.\ \ref{muaver}a for growth 
perpendicular to all 22 planes considered. Thick (green), thin (blue), 
and intermediate (red) solid curves correspond to the same high-symmetry 
planes as in Fig.\ \ref{fbreak}a; long-dashed, short-dashed, and dot-dashed purple 
curves correspond to the same planes with the maximum Miller index of 2 
as in Fig.\ \ref{fbreak}a; thin dark-grey and thick light-grey lines display 
the shear moduli for growth perpendicular to planes with the maximum 
Miller index of 3 and 4, respectively.    

Finally, assuming the {\it polycrystalline} model, i.e. the presence at a 
given depth of crystallites stretched vertically by the same factor
but oriented in all possible ways, we average the effective shear modulus over 
growth orientations. The result is plotted in Fig.\ \ref{muaver}b. The 
family of grey curves shows shear moduli for all 22 planes, the same as
in panel \ref{muaver}a. As in Paper I, we apply two types of 
averaging. For solid (black)
curve, a simple averaging over 22 planes is performed (this result is
denoted below as $\mu_{\rm eff}^{\rm av}$). Dashed 
(orange) curve is obtained by weighted averaging where the weights are 
the numbers of equivalent planes, $N_{\rm eqv}$, given in the first column of Table 
\ref{latparam}. Both types of averaging produce very similar curves. 
The same effect was observed in Paper I for the bc lattice. The fit
for simple-averaged shear modulus of the bc lattice from Paper I 
(equation 20) is 
reproduced here by dot-dashed (dark-green) line for comparison. Solid 
(red) curve is the effective shear modulus, $\mu_{\rm eff}^{\{111\}}$, for crystallites 
growing orthogonal to $\{111\}$ planes, which plausibly should be used 
in the {\it macrocrystallite} model. To facilitate further usage, we have 
fitted $\mu_{\rm eff}^{\rm av}(\xi)$ and $\mu_{\rm eff}^{\{111\}}(\xi)$
for the fc lattice (to rms fit accuracy better than 0.01\%) by analytic formulae:
\begin{eqnarray}
    \mu_{\rm eff}^{\rm av} &=& -0.18835+0.3\,\xi^{1/3}~,
\\
    \mu_{\rm eff}^{\{111\}} &=& 0.28857-0.713\,\xi^{-1/3}+0.5\,\xi^{-1}~.
\label{mufit}
\end{eqnarray}

\section{Discussion}
The results for breaking shear strain for the fc lattice are very similar to those for the bc lattice obtained in
Paper I. 
Shear deformations in different planes are characterized by significantly 
different breaking properties.     
A common property for all planes (except for $\{100\}$ planes)
is a reduction of  
breaking shear strain with deviation of the stretch factor from 1 and 
especially with its approach to the breaking limit for stretch. 
Just like 
in the case of the bc lattice,
the $\varepsilon_{\rm crit}^{\rm min}(\xi)$ curve in the {\it polycrystalline} model is essentially 
parabolic. 
The locus of breaking strain minima of the fc
lattice lies systematically below that for the bc 
structure. We do not think that this is related to the fact that fewer 
stretch orientations have been analyzed in Paper I. 

For the unstretched fcc lattice, 
$\varepsilon_{\rm crit}^{\rm min}(1) \approx 0.055$ (to be compared with
$\approx 0.064$ for bcc). If we artificially reduce this number by $\sim 25\%$ to
account for lattice imperfections \citep[following][]{HK09}, we obtain
$\sim 0.04$ ($\sim 0.05$ for bcc). Breaking stretch factor for the
fc lattice is found as $\xi_{\rm crit}\approx 0.95$ and 1.06 for contractions
and elongations, respectively. For breaking stretch strain, we obtain, respectively,  
$\tilde{\varepsilon}_{\rm crit}\approx -0.034$ and 0.04. This can be
compared with $|\tilde{\varepsilon}_{\rm crit}|\approx 0.04$ reported
for the bc lattice by \citet{BC18}.  

In the case of near-equilibrium bottom-up 
crystallization in a NS crust or a WD core, we expect
growth of fc crystals perpendicular to $\{111\}$ planes in the {\it macrocrystallite} model. Then
the red curve in Fig.\ \ref{fbreak}d rather than the minimum curve would determine the strength
of the crystallized matter. Per equation 15 of Paper I (for NS crust), 
we see that the
critical elongation at the mass density of, for instance, $10^9$ g/cc will be achieved 
for $\sim 1$ m tall crystallites. Since the rightmost segment of the
red curve is almost linear between $\xi \approx 1.04$ and 
$\xi_{\rm crit} \approx 1.08$, there will be an $\sim 5$ cm thick layer
with breaking shear strain $\sim 10$ times smaller than for the bulk of the 
crystallite and an $\sim 5$ mm thick layer with $\sim 100$ times smaller 
breaking shear strain. This can be easily rescaled to other mass 
densities with the key takeaway that $\sim 5\%$ ($\sim 0.5\%$) of 
crystallite height is occupied by layers one (two) orders of magnitude 
weaker than the bulk.   

For the fc lattice, variation of the effective shear modulus
with $\xi$ and with the stretch orientation (i.e. from one shear plane 
to another) is not as strong as for 
the bc lattice (cf.\ figure 4 in Paper I). On the other hand, the averaged 
over stretch directions shear moduli are very similar between the two 
structures. Both types of averaging, simple and weighted, produce very similar curves. 
The same effect was observed in Paper I for the bc lattice.

\section{Conclusion}
It has been argued in Paper I that unidirectional near-equilibrium 
freezing of dense matter in compact stars (from the deeper layers outward) 
preserves the microscopic order of previously crystallized layers and 
is accompanied by gradual vertical stretching of crystal elementary 
cells in response to pressure and electron density reduction. 
Overstretched crystals break, which limits the vertical sizes of 
growing crystallites. In Paper I, elastic and breaking properties of such matter have
been analyzed under the assumption of bc lattice structure. 
In this work, the analysis has been extended to fc 
configurations. In particular, we have studied shear deformations of
stretched fc crystals in planes orthogonal to the stretch 
direction. We have considered shear in all 22 crystallographic planes 
characterized by Miller indices $\leq 4$. This should be contrasted with
the analysis in Paper I, where only 7 planes (3 of the highest
symmetry and 4 less symmetric) have been studied.

Our conclusions are very similar to those of Paper I. 
In particular, a common property for all planes is a reduction of 
breaking shear strain with deviation of the stretch factor from 1 and 
especially with its approach to the breaking limit for stretch. Shear 
deformations in different planes are characterized 
by significantly different breaking properties. Assuming the 
{\it polycrystalline} model of matter, the breaking shear strain at each stretch 
factor $\xi$ has been minimized over stretch directions. Similar to 
the bc configuration, the locus of these minima was found to be 
essentially parabolic. At $\xi=1$, the minimum breaking shear parameter
$\varepsilon_{\rm crit}^{\rm min}(1) \approx 0.055$ (to be compared with
$\approx 0.064$ for bcc). The $\varepsilon_{\rm crit}^{\rm min}(\xi)$ 
curve for the fc structure lies systematically below that for the bc 
structure. We do not think that this is related to the fact that fewer 
stretch orientations have been analyzed in Paper I. In both cases, for 
certain $\xi$, the minima are achieved for low-index planes. For 
convenience, the $\varepsilon_{\rm crit}^{\rm min}(\xi)$ curve for the 
fc lattice has been fitted by a simple analytic expression.  

We have also determined the effective shear moduli vs. the stretch 
factor for shear deformations perpendicular to the stretch direction for 
all 22 shear planes. For the fc lattice, the variation of these curves
with $\xi$ and from one shear plane to another is not as strong as for 
the bc lattice (cf.\ figure 4 in Paper I). On the other hand, the averaged 
over stretch orientations shear moduli are very similar between the two 
structures. 

It is possible that the epitaxial crystallization in 
compact star matter results in a formation of large-scale crystallites. 
Optionally, it seems plausible that, 
in agreement with the Bravais's rule, the entire crystallization front 
grows orthogonal to the close-packed planes, which, for the fc 
structure, are the $\{111\}$ planes (see Paper I for discussion). In 
this {\it macrocrystallite} scenario,
peculiar properties of ideal crystals oriented in a certain way may 
manifest themselves in astrophysical phenomena. For instance, if we 
assume the fc lattice growth perpendicular to the $\{111\}$ planes, 
we expect the appearance of layers with the thickness of $\sim 5\%$ 
($\sim 0.5\%$) of the crystallite height whose breaking shear strain 
is 10 (100) times lower than for the bulk of the crystallite. 
Such structure of stellar
matter may help explain rich observational phenomenology of magnetars 
with active episodes accompanied by numerous bursts and outbursts.

\end{document}